\documentclass[12pt,preprint]{aastex}

\newcommand{\bdv}[1]{\mbox{\boldmath$#1$}}

\def\au{{\rm AU}}

\def\masyr{{\rm mas}\,{\rm yr}^{-1}}

\def\mas{{\rm mas}}

\def\muas{\mu{\rm as}}

\def\rel{{\rm rel}}

\def\e{{\rm E}}

\def\bmu{{\bdv\mu}}

\begin{document}
\title{KMT-2016-BLG-1107: A New Hollywood-Planet Close/Wide Degeneracy}

 \author{\textsc{
Kyu-Ha Hwang$^{1}$, 
Yoon-Hyun Ryu$^{1}$, 
Hyoun-Woo Kim$^{1}$, 
Michael D.Albrow$^{2}$, 
Sun-Ju Chung$^{1,3}$, 
Andrew Gould$^{1,4,5}$, 
Cheongho Han$^{6}$, 
Youn Kil Jung$^{1}$, 
In-Gu Shin$^{7}$, 
Yossi Shvartzvald$^{8,9}$, 
Jennifer C. Yee$^{7}$, 
Weicheng Zang$^{10}$,
Sang-Mok Cha$^{1,11}$, 
Dong-Jin Kim$^{1}$,
Seung-Lee Kim$^{1,3}$, 
Chung-Uk Lee$^{1,3}$, 
Dong-Joo Lee$^{1}$,
Yongseok Lee$^{1,11}$, 
Byeong-Gon Park$^{1,3}$, 
Richard W. Pogge$^{5}$} }

\affil{$^{1}$Korea Astronomy and Space Science Institute, Daejon
34055, Korea}

\affil{$^{2}$University of Canterbury, Department of Physics and
Astronomy, Private Bag 4800, Christchurch 8020, New Zealand}

\affil{$^{3}$Korea University of Science and Technology, Daejeon 34113, Korea}

\affil{$^{4}$Max-Planck-Institute for Astronomy, K\"{o}nigstuhl 17,
69117 Heidelberg, Germany}

\affil{$^{5}$Department of Astronomy, Ohio State University, 140 W.
18th Ave., Columbus, OH 43210, USA}

\affil{$^{6}$Department of Physics, Chungbuk National University,
Cheongju 28644, Republic of Korea}

\affil{$^{7}$ Harvard-Smithsonian Center for Astrophysics, 60 Garden
St., Cambridge, MA 02138, USA}


\affil{$^{8}$IPAC, Mail Code 100-22, Caltech, 1200 E. California Blvd., 
Pasadena, CA 91125, USA}

\affil{$^{9}$NASA Postdoctoral Program Fellow}

\affil{$^{10}$Physics Department and Tsinghua Centre for
Astrophysics, Tsinghua University, Beijing 100084, China}



\affil{$^{11}$School of Space Research, Kyung Hee University,
Yongin, Kyeonggi 17104, Korea}

\begin{abstract}

We show that microlensing event KMT-2016-BLG-1107 displays a new type of degeneracy between
wide-binary and close-binary Hollywood events in which a giant-star source
envelops the planetary caustic.  The planetary anomaly takes the form
of a smooth, two-day ``bump'' far out on the falling wing of the light curve,
which can be interpreted either as the source completely enveloping a 
minor-image caustic due to a close companion with mass ratio $q=0.036$,
or partially enveloping a major-image caustic due to a wide companion with
$q=0.004$.  The best estimates of the companion masses are both in the
planetary regime ($3.3^{+3.5}_{-1.8}\,M_{\rm jup}$ and 
$0.090^{+0.096}_{-0.037}\,M_{\rm jup}$) but differ by an even larger factor
than the mass ratios due to different inferred host masses.
We show that the two solutions can be distinguished by
high-resolution imaging at first light on next-generation (``30m'')
telescopes.  We provide analytic guidance to understand the conditions
under which this new type of degeneracy can appear.

\end{abstract}


\section{{Introduction}
\label{sec:intro}}

\citet{gould97} proposed a ``Hollywood'' strategy of searching for
microlensing planets by ``following the big stars'', which made 
use of the fact that planets normally betray their presence in microlensing
events when the source star passes over, or very close to, a caustic
generated by the planet.  For low-mass planets (hence, small caustics),
the probability for the source to pass over the caustic is therefore
proportional to the size of the source, rather than of the much smaller
caustic.  \citet{gg97} showed that for ``wide'' ($s>1$)
topologies, the excess magnification $\Delta A$ (relative to the
case of a point lens without planetary companions) when the source
fully envelops the caustic is 
\begin{equation}
\Delta A \rightarrow {2 q\over\rho^2} = {2 m_p\theta_\e^2\over M\theta_*^2},
\label{eqn:deltaA}
\end{equation}
i.e., exactly the same value as would be the case if the planet were
isolated.  Here, $q=m_p/M$ is the planet-host mass ratio, $s$ is the
planet-host separation normalized to the Einstein radius $\theta_\e$,
$\rho=\theta_*/\theta_\e$, $\theta_*$ is the angular source radius,
\begin{equation}
\theta_\e = \sqrt{\kappa M\pi_\rel};
\qquad
\kappa\equiv {4 G\over c^2\au}\simeq 8.14\,{\mas\over M_\odot},
\label{eqn:thetae}
\end{equation}
and $\pi_\rel=\au(D_L^{-1}-D_S^{-1})$ is the lens-source relative parallax.
If we adopt a threshold of detectability of, e.g., $\Delta A\ga 0.1$, then
we can rewrite Equation~(\ref{eqn:deltaA}) as
\begin{equation}
m_p = {\theta_*^2\Delta A\over 2\,\kappa\pi_\rel}
\ga 0.75 M_\oplus\biggl({\theta_*\over 6\,\muas}\biggr)^2
\biggl({\pi_\rel\over 0.1\,\mas}\biggr)^{-1},
\label{eqn:deltaA2}
\end{equation}
where we have normalized to the angular source size of a typical
clump giant and to the relative parallax of a typical disk lens.
Hence, this approach is potentially sensitive to very low-mass
planets, particularly because large stars are also bright
(unless they are heavily extincted), meaning that the photometric
precision is generally good.

The virtue of this approach was first illustrated by OGLE-2005-BLG-390,
which was intensively monitored by the PLANET collaboration, leading
to the detection of a $q\sim 8\times 10^{-5}$ planet, with estimated
mass $m_p\sim 5\,M_\oplus$ \citep{ob05390}.  This was only the third
published planet, and still to this date, one of only seven microlensing
planets with well-measured mass ratios in the range $q<10^{-4}$
\citep{ob171434}.

Nevertheless, this channel has been the subject of remarkably little
systematic study.  The first fundamentally new development was the
discovery of a potential degeneracy between ``Cannae'' and ``von Schlieffen''
Hollywood events, in which the source, respectively, fully and partially
envelops the caustic \citep{ob170173}.

Here we analyze the Hollywood microlensing event KMT-2016-BLG-1107 and
report the discovery a second potential (``close/wide'') degeneracy.
Although this degeneracy has a similar $s\leftrightarrow s^{-1}$
symmetry to the well-established ``close/wide'' degeneracy that affects 
central caustics \citep{griest98,dominik99}, the two degeneracies are
not fundamentally related.  We study the general conditions under
which this close/wide planetary-caustic degeneracy can arise and show
that (in strong contrast to the close/wide central-caustic degeneracy),
the mass ratios $q$ of the two degenerate solutions generically differ
by a relatively large factor, with the $s<1$ solution having a more
massive planet.  In the present case, $q=4.0\times 10^{-3}$ and
$q=3.6\times 10^{-2}$ for the $s>1$ and $s<1$ solutions, respectively.  
Finally, we show that imaging
with adaptive optics (AO) cameras will be able to resolve the degeneracy
at first light on next-generation (``30m'') telescopes.

\section{{Observations}
\label{sec:obs}}

KMT-2016-BLG-1107 is at (RA,Dec) = (17:45:40.26,$-26$:01:54.48)
corresponding to $(l,b)=(2.5,1.5)$.  
It was discovered by applying the Korea Microlensing Telescope Network
(KMTNet, \citealt{kmtnet}) 
post-season event finder \citep{eventfinder} to 2016 KMTNet data
\citep{2016eventfinder}.  These data were taken on KMTNet's
three identical 1.6m telescopes at CTIO (Chile, KMTC), 
SAAO (South Africa, KMTS) and SSO (Australia, KMTA), each equipped
with identical $4\,{\rm deg}^2$ cameras.  The event lies in KMTNet field
BLG18 with a nominal cadence of $\Gamma = 1\,{\rm hr^{-1}}$.  
In fact, the cadence was altered from April 23 to June 16
($7501 < {\rm HJD}^\prime\equiv {\rm HJD}-2450000<7555$) to support
the {\it Kepler K2} C9 microlensing campaign 
\citep{gouldhorne,henderson16,2016k2}.  During this period,
the cadence was reduced to $\Gamma = 0.75\,{\rm hr^{-1}}$  for
KMTS and KMTA, but remained at $\Gamma = 1\,{\rm hr^{-1}}$  for KMTC.
We note that KMTC data are affected by a bad column on the CCD and so often
have significantly larger error bars than KMTS and KMTA data.

The great majority of observations were carried out in the $I$ band 
with occasional $V$-band observations made
solely to determine source colors.
All reductions for the light curve
analysis were conducted using the pySIS implementation \citep{albrow09}
of difference image analysis (DIA,
\citealt{alard98}). While the $V$-band data are sufficient to measure the color (Section~\ref{sec:thetae}), because the source is very red and there are many fewer $V$-band points than $I$-band, the $V$-band data do not place significant constraints on the modeling. Thus, we do not use them in the modeling.

\subsection{{Removal of Long-Term Trend}
\label{sec:trend}}

\begin{figure}
\plotone{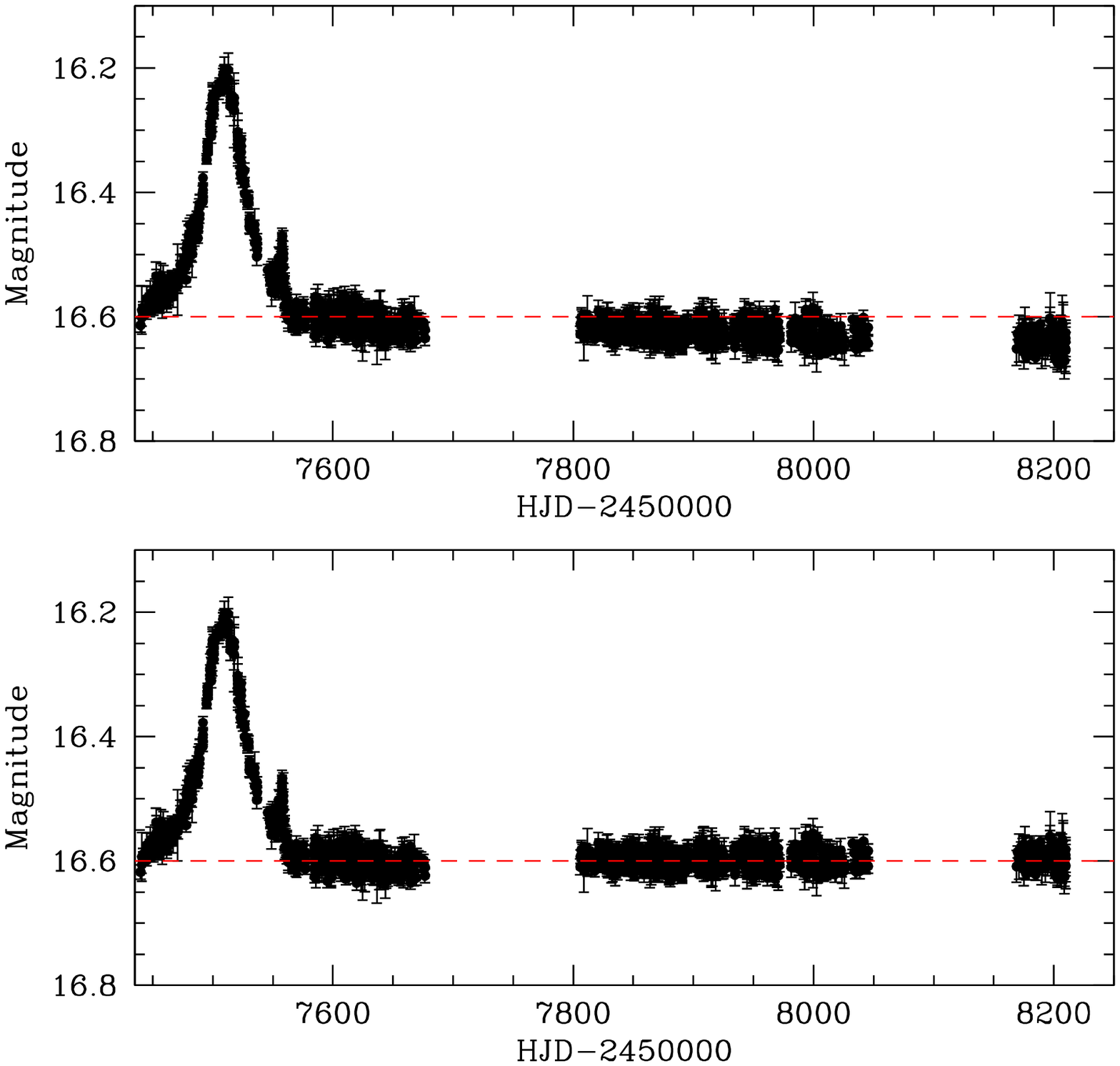}
\caption{KMTC light curve of KMT-2016-BLG-1107 before (upper panel) and
after (lower panel) removal of a linear trend in the baseline flux,
which corresponds to $d I_{\rm base}/dt = +0.018\,{\rm mag\,yr^{-1}}$.  
The fit was done to 2017 data, while the extension to 2018
serves as a check.  The slopes found for KMTA and KMTS are nearly
identical.
}
\label{fig:trend}
\end{figure}

The raw light curve shows a long-term trend in the combined 2016-2018
data.  See Figure~\ref{fig:trend}.  The origin of this trend is unknown.
It could, for example, be due to blended light from a relatively 
high-proper-motion nearby star.  Whatever the cause, it is almost
certainly unrelated to the primary microlensing event or the additional
``bump'' in the lightcurve.  We therefore fit the baseline of the 
light curve to a constant plus a slope and remove the slope
(see Figure~\ref{fig:trend}) before undertaking the microlensing
analysis.

\section{{Light Curve Analysis}
\label{sec:anal}}

The resulting light curve of KMT-2016-BLG-1107 in the neighborhood of
the event is shown in Figure~\ref{fig:lc}.
It primarily takes the form of a standard \citet{pac86} 
single-lens/single-source (1L1S) curve, which is characterized by
three geometric parameters $(t_0,u_0,t_\e)$.  These are, respectively,
the time of maximum, the impact parameter (normalized to $\theta_\e$),
and the Einstein timescale,
\begin{equation}
t_\e \equiv {\theta_\e\over\mu_\rel},
\label{eqn:tedef}
\end{equation}
where $\bmu_\rel$ is the lens-source relative proper motion.

However, in addition there is a small, short-lived ``bump'' 
on the falling wing of the light curve at ${\rm HJD}^\prime\simeq 7557$.
This appearance is qualitatively similar to the classic Hollywood
event, OGLE-2005-BLG-390 \citep{ob05390}, and so plausibly could
be generated by a similar major-image ($s>1$) caustic that is fully
enveloped by the source.  To test this conjecture, we conduct a
systematic grid search over the seven standard parameters of 
binary-lens/single-source (2L1S) events.  There are the three \citet{pac86}
parameters just mentioned $(t_0,u_0,t_\e)$, the three parameters mentioned
in Section~\ref{sec:intro}, $(s,q,\rho)$, and the angle $\alpha$ between
the binary axis and the direction of lens-source relative motion $\bmu_\rel$.
In addition to these geometric parameters, there are two flux parameters
$(f_s,f_b)$ for each observatory, $i$, so that the observed fluxes $F_i(t)$ are
modeled by 
\begin{equation}
F_i(t) = f_{s,i} A(t;t_0,u_0,t_\e,\rho,s,q,\alpha) + f_{b,i} .
\label{eqn:foft}
\end{equation}

\subsection{{Two Solutions:  Wide and Close Planetary Companions}
\label{sec:twosol}}

We employ Monte Carlo Markov Chain (MCMC) $\chi^2$ minimization to
carry out a grid search, in
which $(t_0,u_0,t_\e,\rho,\alpha)$ are allowed to vary, while $(s,q)$
are held fixed.  The chains are seeded with $(t_0,u_0,t_\e)$ derived
from an initial \citet{pac86} fit and $\rho=0.04$ based on the relative
duration of the two bumps.  For each $(s,q)$, $\alpha$ is seeded at
six values that are spaced uniformly around the unit circle.

\begin{figure}
\plotone{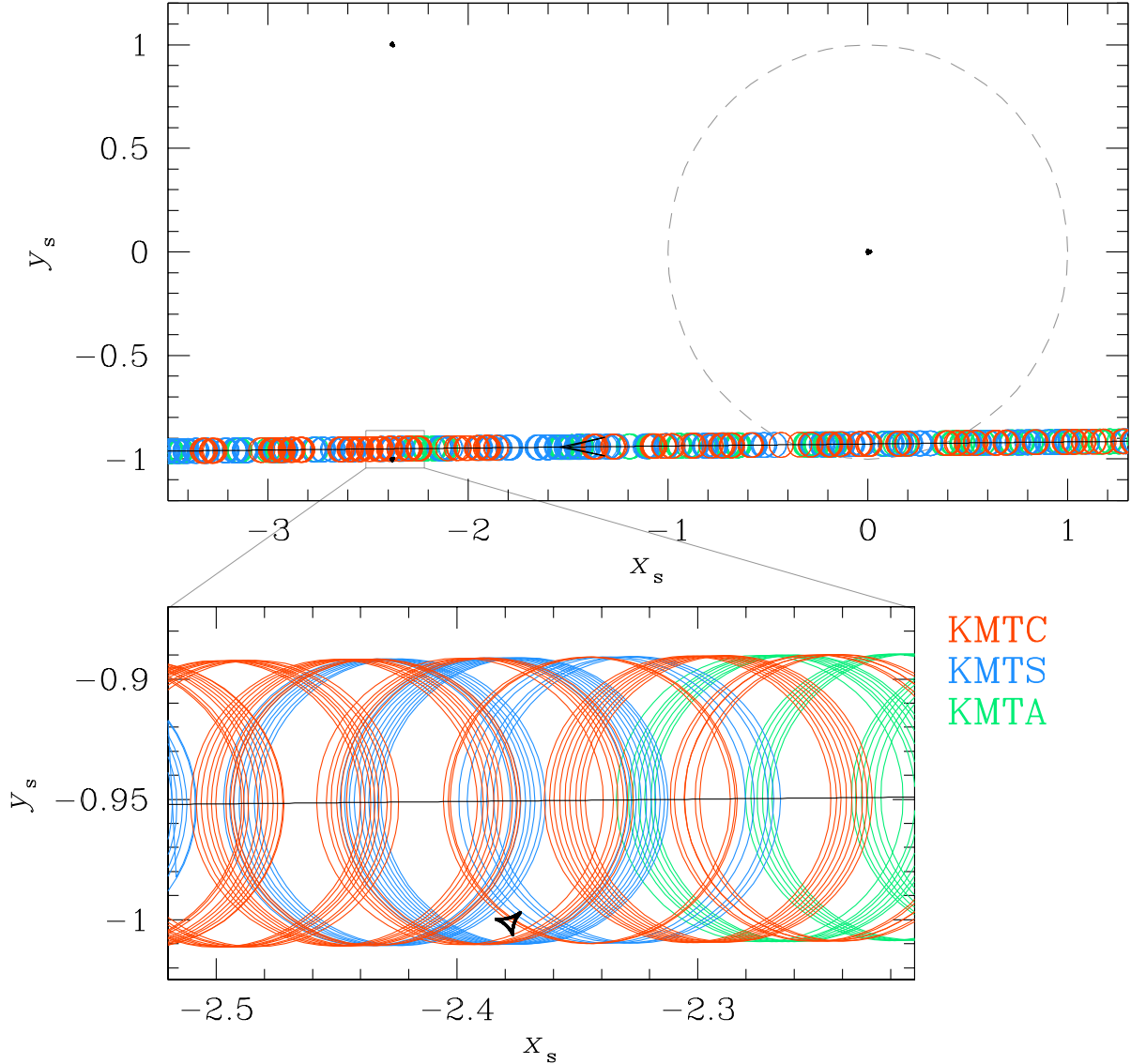}
\caption{Caustic geometry of KMT-2016-BLG-1107,
close $(s<1)$ solution.  The upper panel
shows the source trajectory, which grazes the Einstein ring
(dashed line) and then fully envelops the tiny triangular caustic,
roughly 2.5 Einstein timescales $t_\e$ later.  This Cannae-type
Hollywood envelopment is shown in greater detail in the lower-panel
zoom, where the source position is shown by circles at the epochs
of observations from the three color-coded observatories.
During the time that the caustic passes close to the limb, it spends
more than half a source-diameter-crossing time inside the source.
This should be compared to Figure~\ref{fig:geomw}.
}
\label{fig:geomc}
\end{figure}

\begin{figure}
\plotone{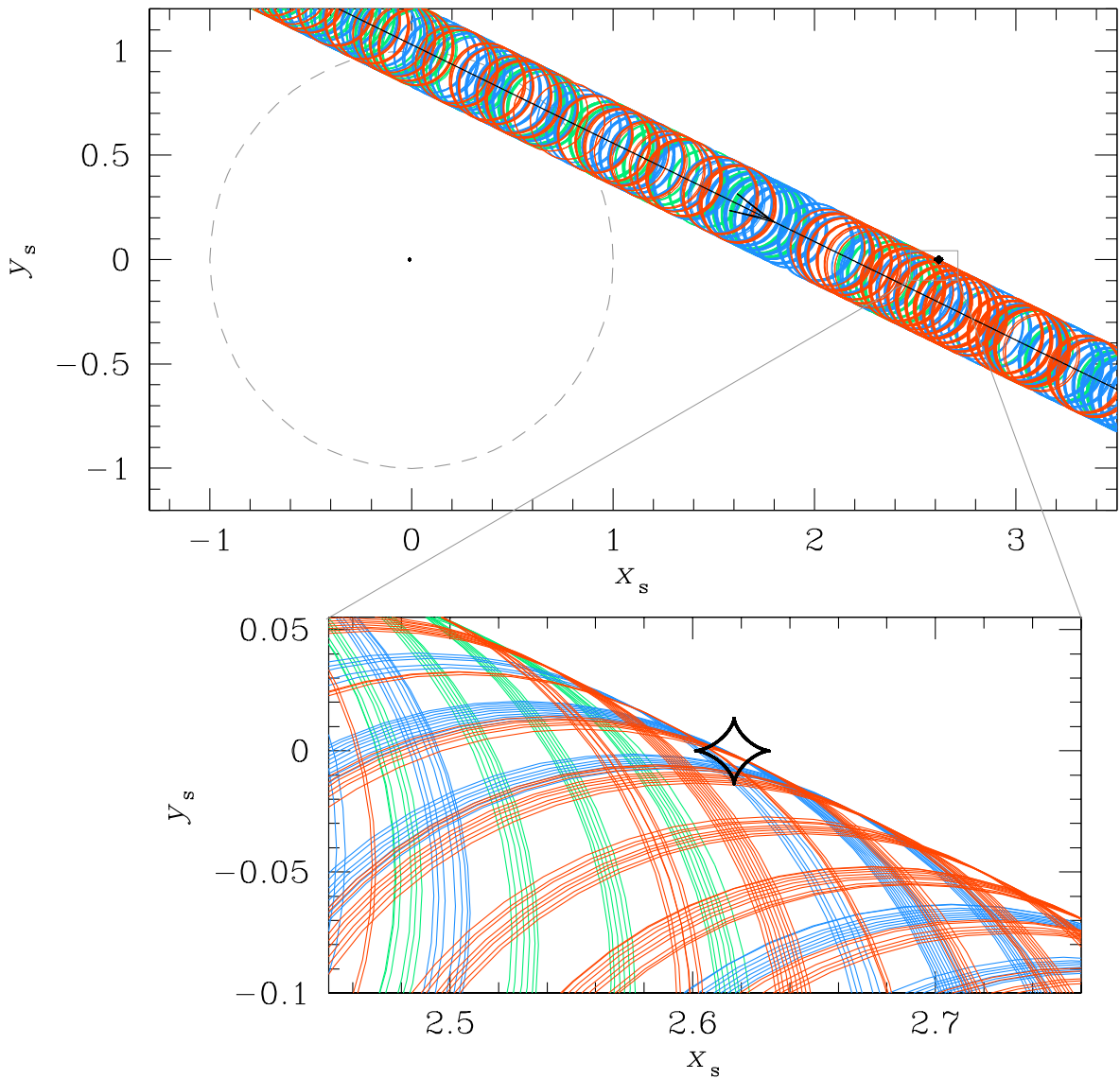}
\caption{Caustic geometry of KMT-2016-BLG-1107,
wide $(s>1)$ solution.  The upper panel
shows the source trajectory, which grazes the Einstein ring
(dashed line) and then partially envelops the small quadrilateral
planetary  caustic,
roughly 2.5 Einstein timescales $t_\e$ later.  This von Schlieffen-type
Hollywood envelopment is shown in greater detail in the lower-panel
zoom, where the source position is shown by circles at the epochs
of observations from the three color-coded observatories.  
This grazing geometry implies that the normalized source size $\rho$
must be substantially larger than for the close geometry (compare
to Figure~\ref{fig:geomc}) in order to generate a ``bump'' of similar
duration.  This implies a substantially lower proper motion $\mu_\rel$,
which will eventually permit future adaptive optics (AO) observations
to distinguish between the two solutions.  
See Section~\ref{sec:thetae} and, in particular, 
Equations~(\ref{eqn:thetaemussmall}) and (\ref{eqn:thetaemusbig}).
}
\label{fig:geomw}
\end{figure}

The grid search yields only two minima whose geometries are shown in Figures~\ref{fig:geomc} and \ref{fig:geomw}. In both cases, the source passes over a small planetary caustic that lies far from the lens. For the solution with $s < 1$, there are two such caustics, so we also check for a solution that crosses over the caustic at $(x_{\rm s}, y_{\rm s}) \sim (-2.4, 1.0)$. This solution is disfavored by $\Delta\chi^2 \sim 140$ of which $\Delta\chi^2\sim 50$ comes from the rising side of the light curve, when the source is forced to cross the "de-magnified zone" even though no such demagnification is seen in the light curve. Furthermore, this solution has severe negative blending that makes it unphysical.

We further refine the two good solutions
with additional MCMC runs, in which all seven parameters are
allowed to vary.  
These two solutions are shown in Table~\ref{tab:ulens}. The flux values are quoted for the KMTC dataset.
We note first that the \citet{pac86} parameters $(t_0,u_0,t_\e)$
are essentially identical between the two solutions.  This is consistent
with the fact that the light curve morphology is dominated by a
broad bump that is generated by the host.  Similarly, the flux
parameters $(f_s, f_b)$ are also essentially identical.\footnote{The flux system is defined so that $I = 18 - 2.5\log_{10}f_x$.}

Next we note that the two values of $s$, (0.345 and 2.96), almost perfectly obey
$s\leftrightarrow s^{-1}$.  The \citet{pac86} curve is generated by
two images, which lie at $u_\pm = (u\pm \sqrt{u^2+4})/2$, where $u$
is the projected lens-source separation normalized to $\theta_\e$.
If the planet lies near either image, $s\sim |u_\pm|$, then it
will perturb the image, giving rise to a short-lived deviation.
Because $u_- = -u_+^{-1}$, it follows immediately that planetary
deviations for close and wide solutions should be related by
$s\leftrightarrow s^{-1}$.  

The values of $\alpha$ differ substantially, but this mainly reflects
that the major and minor images (and so any planet perturbing
these images) lie on opposite sides of the host lens.  In the limit
$q\ll 1$, one expects $\alpha\leftrightarrow \pi-\alpha$ from this
effect alone.  However, this symmetry is broken at finite $q$
because the major-image caustic always lies directly on the binary
axis whereas the two minor-image caustics are displaced from the axis
by an angle that increases with $q$ \citep{han06}.  These effects are
apparent from examination of the caustic geometries,
Figures~\ref{fig:geomc} and \ref{fig:geomw}.

The two really important differences between these solutions are in
$q$ and $\rho$.  The mass ratio $q$ is almost 10 times larger
in the close solution while the normalized source size is more than
three times smaller.  Some insight into these differences is
provided by Figures~\ref{fig:geomc} and \ref{fig:geomw}, 
which show the geometries of the two solutions.

In the close solution, the short-lived bump is caused by the
complete (``Cannae'') envelopment of one of the two triangular 
caustics associated with minor-image perturbations, while
in the wide solution it is caused by the partial (``von Schlieffen'')
envelopment of the quadrilateral planetary caustic associated
with major-image perturbations.  In the latter geometry, the
duration of the bump (which is an observed feature of the light
curve) is substantially shorter than the diameter crossing time
of the source, which immediately implies that the normalized source
size must be substantially bigger than for the $s<1$ von Schlieffen solution.

\subsection{{Apparent Preference for Close Solution Is Not Real}
\label{sec:fakepref}}

\begin{figure}
\plotone{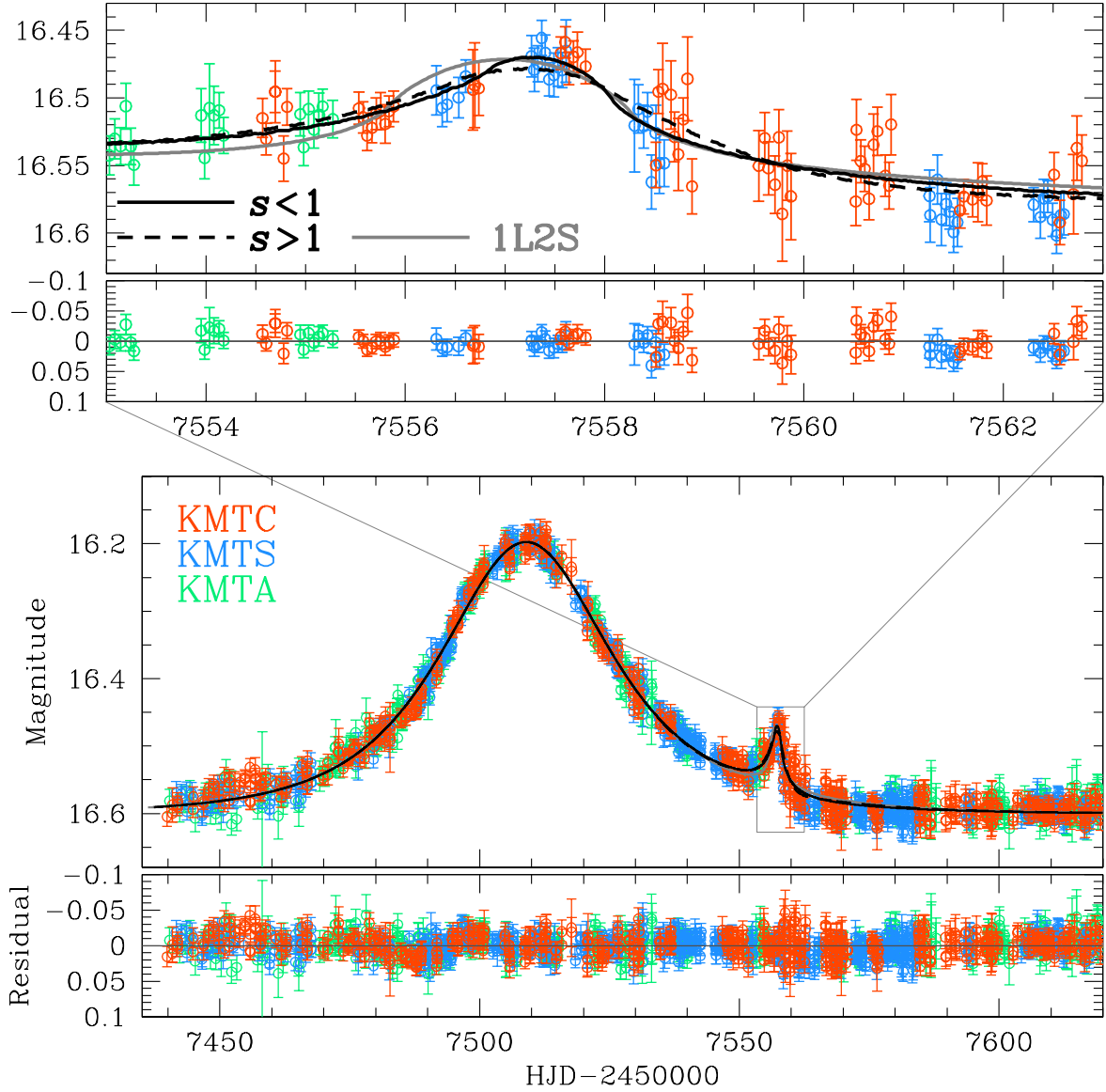}
\caption{KMT-2016-BLG-1107 light curve, together with 
three models: ``close binary lens'' ($s<1$, solid black line),
``wide binary lens'' ($s>1$, dashed line) and ``binary
source'' (1L2S, solid gray line).  The upper panels show a
zoom around the anomaly, while the lower panels show the full event. The residuals are calculated relative
to the $s<1$ model. The 1L2S model is excluded by its failure 
to match the data on the rising wing of the anomaly at 
HJD$^\prime=7556.xx$ as well as its overall high $\chi^2$.
The close $(s<1)$ model is preferred by $\chi^2$ but this
is an artifact of low-level variability, which
depresses the data below either the $s<1$ or $s>1$ models by
several sigma during the interval $7565<{\rm HJD}^\prime<7586$
during which the former model predicts very slightly lower
magnification.
}
\label{fig:lc}
\end{figure}

We note that the close solution is nominally preferred
by $\Delta\chi^2\simeq 40$.  Under most circumstances, we would
regard this as strong evidence in its favor.  However, one
can see from Figure~\ref{fig:lc} that the two models hardly differ,
particularly compared to the error bars, which are relatively large
($\sim 0.006\,$mag for KMTS). We have investigated the origin of this apparently
strong $\chi^2$ difference by constructing the cumulative distribution function
of $\Delta\chi^2=\chi^2_{\rm wide}-\chi^2_{\rm close}$ as a function  of
time, and we find that most of the ``signal'' comes from 
$7565\la {\rm HJD}^\prime\la 7585$, i.e., when $\tau\equiv (t-t_0)/t_\e$
is 3--4 Einstein timescales past peak.  During this interval,
the lightcurve is, on average, several $\sigma$ below either model.
Under such conditions, $\Delta\chi^2\simeq 2\,(\delta/\sigma)((\Delta/\sigma)$
where $\delta$ is the difference between the models, 
$\Delta$ is the difference between the data and the mean of the two
models, and $\sigma$ is the error bar.  That is, for $\Delta/\sigma\sim 2.5$,
we have $\Delta\chi^2=5(\delta/\sigma)$.  In this way, many tens of
points each contribute $\Delta\chi^2\sim 0.5$ even though the precision
of the data does not permit them to distinguish between models.
Inspecting the residuals in Figure~\ref{fig:lc}, we see that they
show irregular variability with an amplitude $\sim 0.02\,$ mag and
a timescale of $\sim 20\,$days.  Hence, we conclude that the
apparent preference of the data for the close solution is an artifact
of this low-level variability, i.e., that a degeneracy persists between 
the close and wide solutions (both are acceptable solutions) despite a difference in chi2.

\subsection{{Binary-Source Solution Excluded}
\label{sec:1L2S}}

Short-lived smooth bumps can be generated by single-lens/binary-source
(1L2S) events \citep{gaudi98}.  In particular, if the secondary source
is substantially fainter than the primary and passes much closer to the lens,
the resulting light curve can mimic a 2L1S planetary event quite well.
We search for binary source solutions, but find that they are disfavored by
$\Delta\chi^2>100$.  See Table~\ref{tab:ulens} in which $q_F, I$ indicates the flux ratio of the two sources.
While some of this signal may come from long-term
variability (see Section~\ref{sec:fakepref}), the 1L2S model clearly
fails to match the data (particular KMTS) on the rising side of
the bump.  See Figure~\ref{fig:lc}.  This short-term failure cannot
be explained by long-term variability.  Hence, we consider that the
1L2S model is excluded.

\section{{Angular Source Radius: $\theta_*$}
\label{sec:thetae}}

The evaluation of the angular source radius $\theta_*$ is generally
important in microlensing events because it leads to the measurement
of the angular Einstein radius and the lens-source relative proper motion
\begin{equation}
\theta_\e = {\theta_*\over\rho};
\qquad
\mu_\rel = {\theta_\e\over t_\e},
\label{eqn:thetastar}
\end{equation}
which can help physically characterize the lens, usually by evaluating
the prior probabilities of these values within the context of a Galactic
model (e.g., \citealt{han95}).

However, in the present case, the evaluation of $\theta_*$ is
of even more fundamental importance.  This is because the two degenerate
solutions identified in Section~\ref{sec:twosol}, which have radically
different mass ratios $q$, also predict radically different proper
motions $\mu_\rel$.  This means that by evaluating $\theta_*$ 
(and so, via Equation~(\ref{eqn:thetastar}), $\mu_\rel$), we can lay
the basis for distinguishing between the two solutions by future
high-resolution imaging.  In particular, we note from Table~\ref{tab:ulens}
that the two solutions have similar source fluxes $f_s$ and similar
Einstein timescales $t_\e$.  As we will see below, similar source
fluxes imply similar $\theta_*$.  Then, from Equation~(\ref{eqn:thetastar}),
the two solutions will be in the relations
\begin{equation}
{\mu_{\rel,s<1}\over \mu_{\rel,s>1}}
\simeq {\theta_{\e,s<1}\over \theta_{\e,s>1}}
\simeq {\rho_{s>1}\over \rho_{s<1}}\simeq 3.09\pm 0.54 .
\label{eqn:murats}
\end{equation}
Hence, the two solutions are very well separated in their {\it relative}
predictions for the proper motion.  The question then is how well the
absolute proper motion can be predicted.  This in turn basically
depends on how well $\theta_*$ can be measured.

We follow the usual approach of measuring the offset of the source
color and magnitude from those of the clump centroid \citep{ob03262}.
The main additional subtlety is that, as discussed in Section~\ref{sec:trend},
the light curve shows a long-term trend in the baseline magnitude
$dI/dt\sim 0.018\pm 0.003\,{\rm mag\,yr^{-1}}$. 

In order to measure the offset of the source from the clump,
we must carry out the lightcurve photometry in a common system with the field
stars.  For this purpose we use pyDIA, i.e., a different package from
pySIS, which is used for the main light-curve analysis.  
 In our initial treatment we apply the slope-removal 
correction derived from the
pySIS analysis to both the $I$-band and $V$-band photometry files
derived from pyDIA.  

In 2016, KMTNet took $V$-band data from both KMTC and KMTS, with relative
$V/I$ cadences of 1/10 and 1/20, respectively.  In addition, as mentioned
in Section~\ref{sec:obs}, the KMTS overall cadence was reduced by 25\% during
most of the event.  Therefore, there are roughly 2.5 times more $V$-band
images from KMTC compared to KMTS over the relevant portions of the light curve.
However, as also mentioned in Section~\ref{sec:obs}, the KMTC data
were adversely affected by a bad column, leading to the loss of many
points and the degradation of some of those remaining.  
Hence, we can expect that both
data sets will contribute roughly equally to the measurement of $\theta_*$ and
so analyze both.

\begin{figure}
\plotone{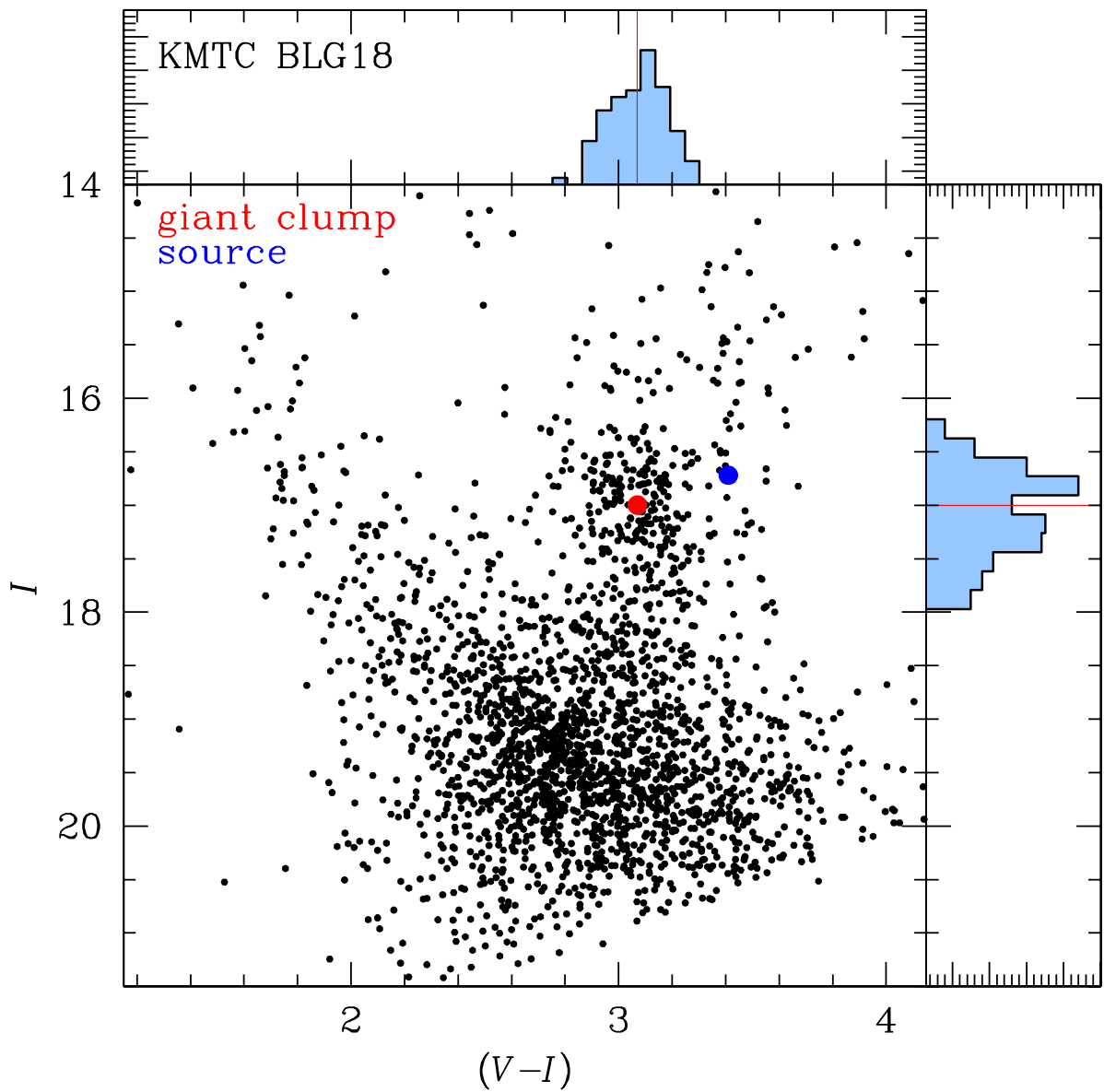}
\caption{Color-magnitude diagram (CMD) based on KMTC data from a
$100^{\prime\prime}$ square centered on KMT-2016-BLG-1107.  The source (blue)
is more than 0.3 mag redder than the centroid of the red clump (red), and
is also more than 0.2 mag brighter.  This offset (confirmed also by
KMTS data -- not shown), leads to a source-radius estimate 
$\theta_*=8.8\pm 0.7\,\mas$.
}
\label{fig:cmd}
\end{figure}

We report first the details of the KMTC measurements and then
summarize those from KMTS.
We find the instrumental $[(V-I),I]$ pyDIA 
positions of the red clump (see Figure~\ref{fig:cmd}),
\begin{equation}
[(V-I),I]_{\rm cl,\rm pyDIA,KMTC}= (2.42,13.97)\pm (0.02,0.04),
\label{eqn:vmiicl}
\end{equation}
and of the source star (by aligning the pyDIA light curves to the best-fitting
model,
\begin{equation}
[(V-I),I]_{s,\rm pyDIA,KMTC}= (2.76,13.72)\pm (0.05,0.01) .
\label{eqn:vmiis}
\end{equation}
The offset is therefore
$\Delta[(V-I),I]_{\rm KMTC}=(0.34,-0.25)\pm(0.06,0.04)$.  
Following the same procedure for KMTS, we obtain
$[(V-I),I]_{\rm cl,\rm pyDIA,KMTS}= (1.60,14.84)\pm (0.02,0.04)$,
$[(V-I),I]_{s,\rm pyDIA,KMTS}= (1.89,14.66)\pm (0.07,0.01)$, and
$\Delta[(V-I),I]_{\rm KMTS}=(0.29,-0.18)\pm(0.07,0.04)$.  
The difference between these two determination has $\chi^2=1.8$
for 2 dof, i.e., perfectly consistent.  We therefore combine the
two measurements to obtain,
\begin{equation}
\Delta[(V-I),I]= (0.32,-0.22)\pm (0.05,0.03) .
\label{eqn:offsetcls}
\end{equation}

We adopt 
$[(V-I),I]_{0,\rm cl}=(1.06,14.36)$ from \citet{bensby13} and 
\citet{nataf13} (for $l=2.5$), and so
\begin{equation}
[(V-I),I]_{0,s}= (1.38,14.14)\pm (0.07,0.15).
\label{eqn:ivmi0s}
\end{equation}
We note that both of the error bars in Equation~(\ref{eqn:ivmi0s})
are larger than those in Equation~(\ref{eqn:offsetcls}).
For the color, we have added in quadrature 0.05 mag as an estimate
of the error in the method, which we derive from the scatter in
the difference between spectroscopic and photometric color estimates
in \citet{bensby13}.  For the magnitude, we add in quadrature a 0.14 error
due to the fractional uncertainty in $f_s$ in Table~\ref{tab:ulens}.

Using the $VIK$ relation
of \citet{bb88} to convert from $V/I$ to $V/K$ and the color/surface-brightness
relation of \citet{kervella04}, we finally derive,
\begin{equation}
\theta_* = 8.83\pm 0.73\,\mu{\rm as} .
\label{eqn:thetastareval}
\end{equation}

The main concern regarding systematic errors in this evaluation is
that the $V$-band data are too noisy to allow us to independently measure
their baseline slope.  More precisely, we find that the slope is consistent
with the $I$-band slope, but with an error that is twice as large as the
value of the $I$-band slope.  Because we do not know the origin of this
slope, it could in principle be substantially different in the two bands.
As a relatively conservative estimate of the
impact of this systematic error, we consider two fairly extreme cases:
first that the $V$-band baseline is flat, i.e., $dV/dt=0$
and second that it is twice the $I$-band slope, i.e., 
$dV/dt=0.036\,{\rm mag\,yr^{-1}}$.  However, we find that even these
changes in assumed slope result in a change in source color of only
0.01 mag, which would result in a change of $\theta_*$ that is more
than an order of magnitude smaller than the error bar in 
Equation~(\ref{eqn:thetastareval}).  Hence, this potential source
of systematic error has no practical impact.  The underlying reason for
this is that we evaluate the source flux only using data from a symmetric
interval around the peak during which the source is significantly 
magnified.  Because the lightcurve is basically symmetric, while the
slope function is anti-symmetric, we expect very little effect.  Indeed,
it is only because the data sampling is higher on the falling wing
of the light curve (because of longer visibility during each night)
that there is any effect at all.

Finally, we estimate $\theta_\e$ and $\mu_\rel$ for the two solutions.
\begin{equation}
\theta_\e = 0.148\pm 0.021\,\mas;
\qquad
\mu_\rel = 2.65\pm 0.38\,\masyr;
\qquad
(s<1),
\label{eqn:thetaemussmall}
\end{equation}
\begin{equation}
\theta_\e = 0.048\pm 0.007\,\mas;
\qquad
\mu_\rel = 0.85\pm0.13\,\masyr;
\qquad
(s>1).
\label{eqn:thetaemusbig}
\end{equation}

Because of the low values of $\mu_\rel$ for both solutions,
resolution of this degeneracy will require the advent of adaptive
optics on next generation (``30m'') telescopes.  To evaluate these
prospects, we begin by recalling 
the experience of \citet{ob05169bat}, who separately resolved the
roughly equally-bright source and lens for OGLE-2005-BLG-169 in $H$-band
($1.65\,\mu$m) using the 10m Keck telescope, at a time when these
were separated by $60\,\mas$, i.e., about 1.5 FWHM.  Considering that
the source is a red giant in the present case, and therefore likely to
be 100-1000 times brighter than the lens, we expect that the minimum
separation is likely 1/3 larger, i.e., 2.0 FWHM.  For the European Extremely
Large Telescope 39m telescope in $J$ band (the most optimistic case),
this would imply a minimum separation of 14 mas.  For an estimated
E-ELT AO first light of 2028, this requirement would be satisfied
by more than a factor 2 for the $(s<1)$ solution, although it would
be only marginally satisfied for the $(s>1)$ solution.  Therefore, there
are reasonable prospects for resolving this degeneracy at AO first light.

\section{{Bayesian Analysis}
\label{sec:bayes}}

Because the planet-host mass ratio $q$ differs by almost a decade
between two solutions that are not distinguishable based on current data,
we cannot give even a relatively precise estimate of the planet mass
based on a Bayesian analysis.  Moreover, if future AO observations
(Section~\ref{sec:thetae}) do distinguish between the two solutions, there
would be no need for a Bayesian analysis because the host mass and distance
will be much better constrained by the measurements of $\theta_\e$ and of the
host flux that derive from those observations.  Nevertheless, for completeness
we carry out a Bayesian analysis to estimate the host mass and distance
for each of the two degenerate solutions.  We employ the same procedures
and Galactic model as did \citet{ob171522}.

\begin{figure}
\plotone{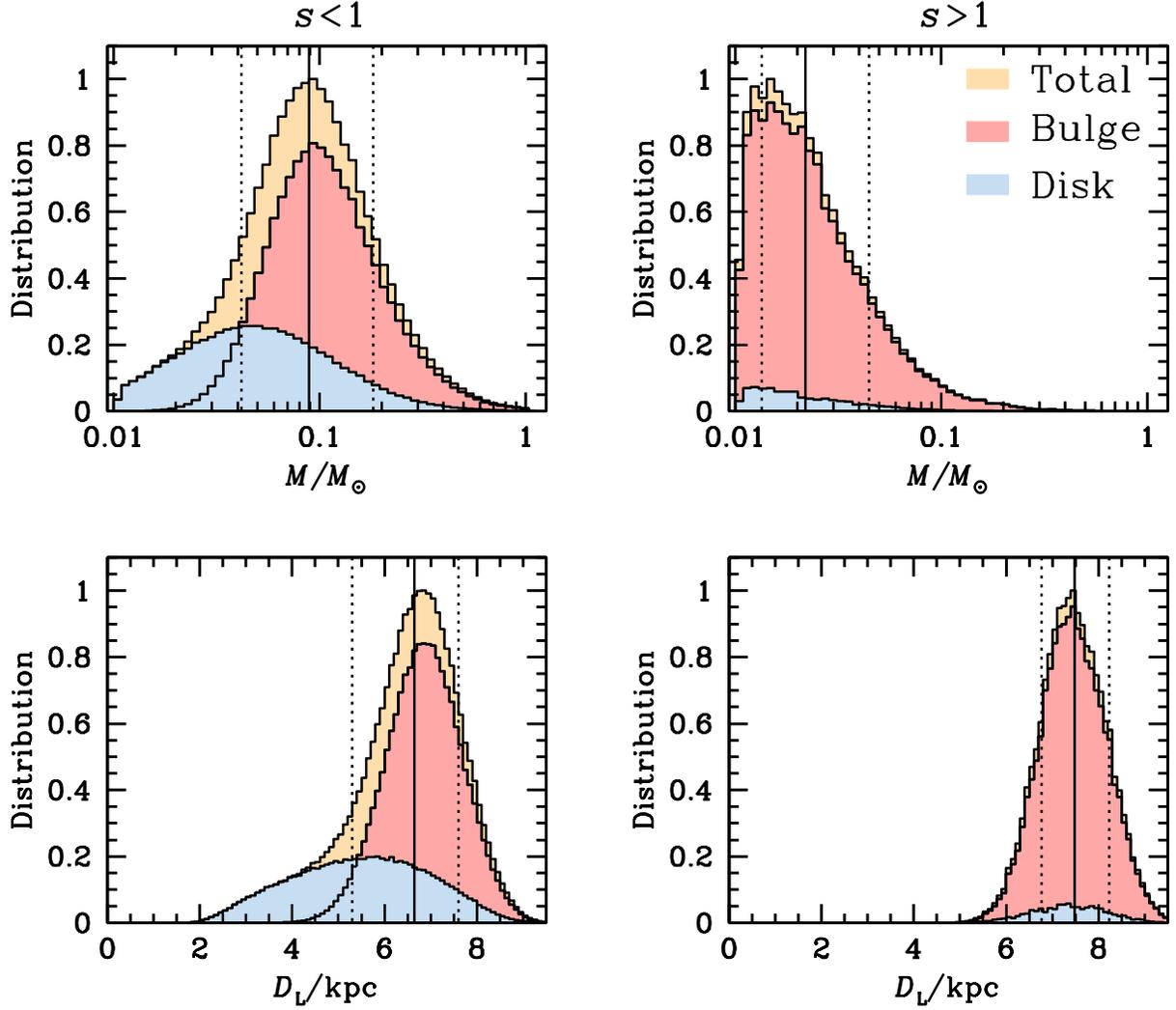}
\caption{Bayesian posteriors for KMT-2016-BLG-1107 for the ``close''
($s<1$, left) and ``wide'' ($s>1$, right) solutions.  For both
solutions, the small Einstein radius ($\theta_\e=0.15\,\mas$ and 
$\theta_\e=0.05\,\mas$) favors a low-mass lens, particularly for the ``wide''
solution.  Similarly, the low proper motion
($\mu_\rel=2.7\,\masyr$ and $\mu_\rel=0.9\,\masyr$) generally favors
a bulge lens, and very strongly so for the ``wide'' solution.
}
\label{fig:bayes}
\end{figure}

The results are illustrated in Figure~\ref{fig:bayes} and summarized
in Table~\ref{tab:phys}.  For both solutions, the small values of
the Einstein radius (Equations~(\ref{eqn:thetaemussmall}) and 
(\ref{eqn:thetaemusbig})) strongly favor low-mass lenses, while the
low proper motions favor Galactic-bulge lenses.  In both cases, the
effect is substantially stronger for the $s>1$ solution.  Also note
that even for the close solution, the planet's projected separation
from its host is relatively large for its mass,
$a_\perp/M\sim 3.9\,\au/M_\odot$.  This would be outside the snowline
for a conventional scaling, $a_{\rm snow} = 2.7\,(M/M_\odot)\,\au$.

According to Figure~\ref{fig:bayes}, there is a roughly 40\% probability
that the lens will be below the hydrogen burning limit (and hence,
likely invisible) even if the $s<1$ solution is correct.  This
means that if future AO observations fail to detect the host, then
we will not be able to determine whether the close or wide solution
is correct: all we will know is that the host is a brown dwarf.

\section{{Discussion: Major/Minor Image Hollywood Degeneracy}
\label{sec:degeneracy}}

\subsection{{Allowed Range of $s$ for Minor-Image Hollywood Events}
\label{sec:allowed_s}}

When \citet{gg97} derived Equation~(\ref{eqn:deltaA}) for the
excess magnification $\Delta A\rightarrow 2q/\rho^2$ generated
by a completely enveloped major-image planetary caustic, they also
derived $\Delta A\rightarrow 0$ for a completely enveloped minor-image 
planetary caustic.  Clearly, this formula fails in the present case,
for which the $s<1$ solution has a ``bump'' that looks qualitatively
similar to classical major-image Hollywood Cannae ``bumps''.  This
is because the \citet{gg97} analytic result implicitly assumed that
the source envelops the {\it entire} minor-image caustic structure, including
both triangular caustics and the trough that lies between them.
In the present case, by contrast, the bump is generated by
passing over one of the two triangular caustics, with the source
well separated from the other triangular caustic and also from
the trough that lies between them.

Hence, the first question to address is: under what conditions can
the source completely envelop one triangular caustic and still be
well separated from the trough.  To address this question, we make
use of the analytic results from \citet{han06}, which are derived in
the limit $q\ll 1$.  \citet{han06} finds that the length of the
planetary caustic (along the direction of the binary axis) is
\begin{equation}
{\Delta\xi_c\over 2} \rightarrow \biggl({27\over 16}\biggr)^{1/2}\sqrt{q} s^3,
\label{eqn:deltaxi}
\end{equation}
while the separation of the center of the caustic from the binary
axis is
\begin{equation}
\eta_{c,0} \simeq
{\sqrt{q}\over s}[(1+s^2)^{-1/2} + (1-s^2)^{1/2}]\rightarrow {\sqrt{q}\over s}
(2 - s^2)
\label{eqn:etac0}
\end{equation}
Hence, the ratio of the distance of the caustic from
the binary axis  to its ``size''  is independent of $q$,
\begin{equation}
R \equiv 
{\eta_{c,0}\over\Delta\xi_c/2}
= \biggl({16\over 27}\biggr)^{1/2}{2 - s^2\over s^4}
\label{eqn:rdef}
\end{equation}
Thus, for $s=(0.4,0.5,0.6,0.7,0.8)$, we have $R=(55,22,10,4.8,2.6)$.
Based on this analysis, we conclude that there can be minor-image
``Hollywood'' type bumps without obvious deviations due to the trough 
only if $s\la 0.7$.

Another way of stating this result is that a symmetric bump at
normalized Einstein-radius position $u_{\rm perturb}\ga 0.7$ can potentially
be explained by either a major-image $(s>1)$ or minor-image $(s<1)$ Hollywood
solution.
On the other hand, for $u_{\rm perturb}\la 0.7$, only major-image Hollywood
solutions are viable.  

In the present case, $s=0.345$ or $s=2.96$, i.e., $u_{\rm perturb}\simeq 2.6$,
we are well into the regime of possible degeneracy between major-image
and minor-image Hollywood events.

We note, however, that while $u_{\rm perturb}\ga 0.7$ is a necessary
condition, it is not
sufficient.  It is also necessary that the perturbation occur on the wing
(rising or falling) of the light curve rather than near the peak.  That is,
if the source crosses one triangular caustic near peak, then it will
also come close to the other one and will cross the pronounced ``dip''
in between.  These features will be easily recognized in the light curve.
In the present case, the anomaly is far into the wing of the light curve
and (in the $s<1$ model) the source trajectory runs nearly parallel 
$(\alpha=\pi+0.11)$ to the binary axis and so essentially
never transits the ``dip''.  See Figure~\ref{fig:geomc}.  Given less than
perfect data, such an extreme geometry is not absolutely required to
avoid detection of the dip, but this consideration does generally
favor larger $u_{\rm perturb}$ to enable an ``effective degeneracy.''

\subsection{{$\Delta a$: Ratio of $q$ for Major-image to Minor-image Solutions}
\label{sec:ratio_q}}

\citet{gg97} gave an analytic formula (Equation~(\ref{eqn:deltaA}))
for the excess magnification $\Delta A$ of Cannae Hollywood events
as a function of $q$ and $\rho$.  For minor-image Hollywood events
there is no such analytic formula, but we can use Equation~(\ref{eqn:deltaA})
as a convenient way to normalize $\Delta A$ to make a numerical study
of this effect.  That is, we define $\Delta a$ as a normalized
$\Delta A$,
\begin{equation}
\Delta a \equiv {\rho^2\over 2q}\Delta A .
\label{eqn:delta_alpha}
\end{equation}

\begin{figure}
\plotone{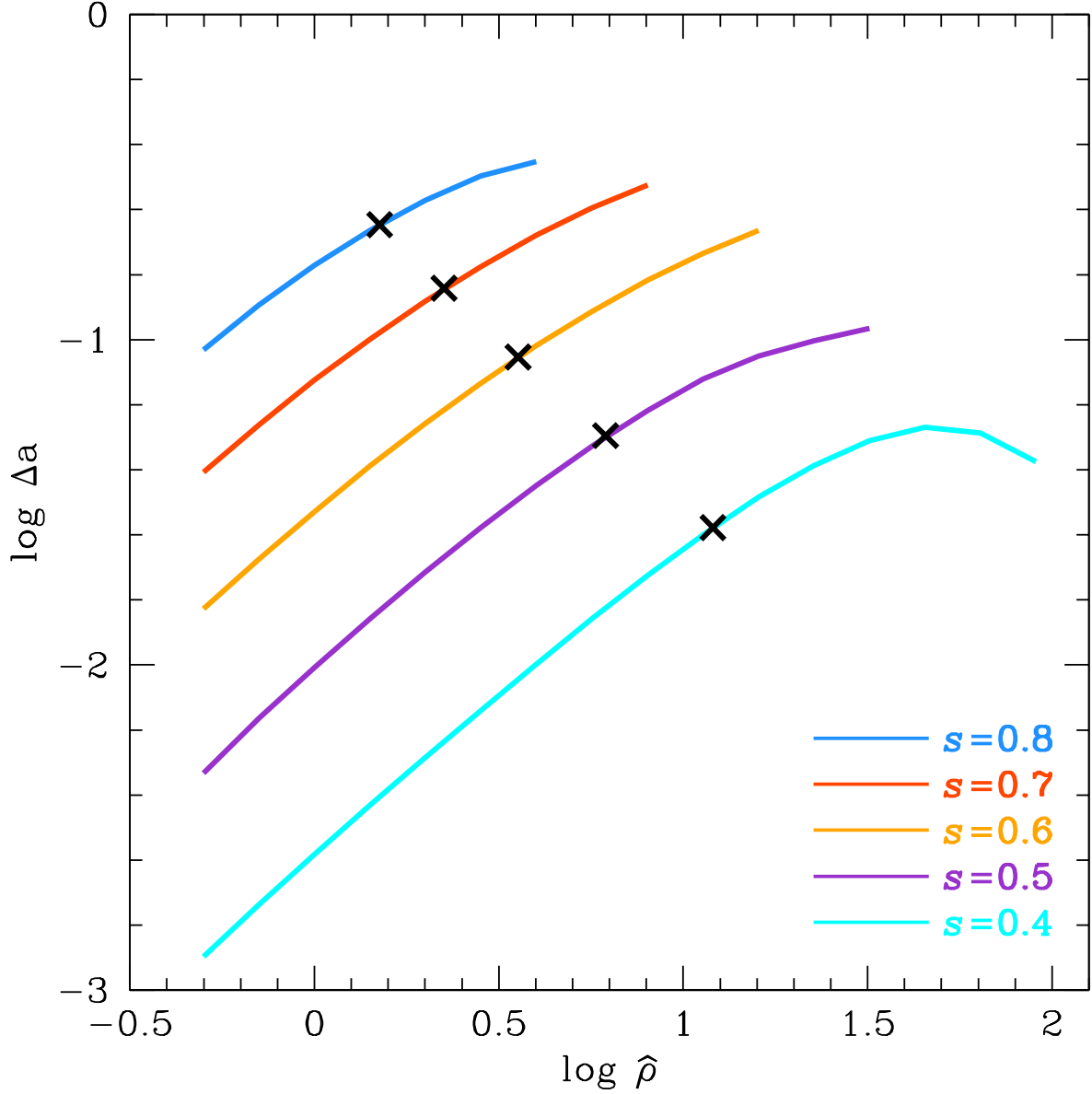}
\caption{Normalized height of minor-image bump $\Delta a$
(scaled to the analytic result from \citealt{gg97} for major-image caustics; 
see Equation~(\ref{eqn:delta_alpha})) versus normalized source size $\hat\rho$
(scaled to caustic size,  see Equation~(\ref{eqn:rhohat})).  This diagram
is for $q=10^{-3}$, but other values of $q$ are similar.  For
each curve, the value of $\hat\rho$ for which $\rho=\sqrt{q}$ is marked
because for major-image caustics, $\rho\ga\sqrt{q}$ implies 
$\Delta a\rightarrow 1$.  The low values of $\Delta a$ generally
imply that envelopment of minor-image caustics require much larger $q$ 
to induce the same size bump compared to major-image caustics.
}
\label{fig:alpha}
\end{figure}

In Figure~\ref{fig:alpha}, we evaluate $\Delta a$ 
for source stars centered
on a minor-image caustic of geometries $(s,q)$, where we fix 
$q=1\times 10^{-3}$ and consider values of $s=(0.4,0.5,0.6,0.7,0.8)$.
In each case, we evaluate $\Delta a$ for values of $\rho$ parametrized
by $\hat\rho$, i.e., $\rho$ normalized to the caustic size:
\begin{equation}
\hat \rho \equiv {\rho\over \Delta\xi_c(s,q)/2} ,
\label{eqn:rhohat}
\end{equation}
where $\Delta\xi_c(s,q)$ is given by Equation~(\ref{eqn:deltaxi}).
In each case, we mark for reference the value of $\hat\rho$ for which 
$\rho=\sqrt{q}$.  This value of $\rho$ is relevant for the comparison
to major-image Hollywood geometries because for the major-image case,
$\rho>\sqrt{q}\Longrightarrow \Delta a\rightarrow 1$.

Figure~\ref{fig:alpha} shows that, particularly for $s\la 0.5$ (i.e., the regime
in which degeneracies are likely to be an issue according to the
analysis of Section~\ref{sec:allowed_s}), $\Delta a$ never rises
much above 0.1, i.e., at least an order of magnitude smaller than
for the generic major-image case.  Hence, according to 
Equation~(\ref{eqn:delta_alpha}), and assuming similar values of $\rho$,
much larger values of $q$ are required to generate a given amplitude
of ``bump'' for minor-image Cannae envelopment compared to major-image
Cannae envelopment.

As we have seen for the case of KMT-2016-BLG-1107, the inferred
values of $\rho$ are not in fact necessarily the same for the
major-image and minor-image solutions.  The two solutions must produce
the same duration ``bump'', but they might achieve this by the caustic
traversing different chords through the source, or (as in the present case)
by one solution being Cannae and the other von Schlieffen.  Nevertheless,
Figure~\ref{fig:alpha} indicates an overall tendency toward higher
$q$ for the close $(s<1)$ solution, a tendency that is in fact realized
in the present case.

\subsection{{Future Prospects for the Major/Minor Image Hollywood Degeneracy}
\label{sec:future}}

The short, smooth bump experienced by KMT-2016-BLG-1107 2.5 Einstein
crossing times after peak makes it similar in some respects to
OGLE-2008-BLG-092 \citep{ob08092}, MOA-2012-BLG-006 \citep{mb12006}, and MOA-2013-BLG-605 \citep{mb13605}.
These had $u_{\rm perturb}=4.87$, $u_{\rm perturb}=4.18$, and $u_{\rm perturb}=1.91$ respectively. That is the first two were even further into the wing of the light curve than KMT-2016-BLG-1107, while the last had a comparable value.
However, in contrast to KMT-2016-BLG-1107, all three of these earlier
events were unambiguously
interpreted as having major-image perturbations and, therefore, very wide
companions.  For MOA-2012-BLG-006 and MOA-2013-BLG-605, this was facilitated by the relatively
small source size ($\rho/\sqrt{q}=0.09$ and 0.2, respectively), of the same order as the caustic,
which induced some structure on the planetary bump.  As noted by
\citet{ob08092,mb12006}, events of this type provide a unique probe
of very wide separation planets (``Uranus'' and ``Neptune'' analogs),
provided of course that they can be unambiguously interpreted as
due to very wide (rather than very close) separation companions.

The KMTNet survey is well-suited to detect more events of this type.
The radius crossing time of 
typical giant sources with $\theta_*\sim 6\,\muas$, is
$t_* = 13\,{\rm hr}(\mu_\rel/[4\,\masyr])$.  Thus, in the $41\,{\rm deg}^2$
that are observed at cadences $\Gamma\geq 1\,{\rm hr}^{-1}$ (like BLG18),
Hollywood perturbations should be covered very well (weather permitting).
Even the additional $44\,{\rm deg}^2$ that are observed at
$\Gamma\geq 0.4\,{\rm hr}^{-1}$ should yield adequate coverage in
many cases.  However, more practical experience will be required
to determine how often such detections are impacted by degeneracies,
and under what conditions these can be resolved.


\begin{deluxetable}{lccc}
\tablecolumns{4} \tablewidth{0pc}\tablecaption{\textsc{Best-fit
solutions}} \tablehead{ \colhead{Parameters } & \colhead{$s<1$} &
\colhead{$s>1$} & \colhead{1L2S}} \startdata
  $\chi^2/\rm{dof}$               &3894.913/3895         &3936.813/3895        &4045.87/3895        \\
  $t_0$ $(\rm{HJD}^{\prime})$     &7508.681 $\pm$ 0.068  &7509.227 $\pm$ 0.057 &7509.034 $\pm$ 0.054 \\
  $u_0$                           &0.927 $\pm$ 0.061     &0.932 $\pm$ 0.057    &0.913 $\pm$ 0.068    \\
  $t_{\rm E}$ $(\rm{days})$       &20.403 $\pm$ 0.875    &20.380 $\pm$ 0.805   &20.818 $\pm$ 0.974   \\
  $s$                             &0.345 $\pm$ 0.013     &2.965 $\pm$ 0.104    &...                  \\
  $q$  $(10^{-2})$                &3.614 $\pm$ 0.544     &0.398 $\pm$ 0.059    &...                  \\
  $\alpha$ $(\rm{rad})$           &3.132 $\pm$ 0.025     &0.441 $\pm$ 0.012    &...                  \\
  $\rho$ $(10^{-2})$              &5.955 $\pm$ 0.683     &18.411 $\pm$ 2.423   &...                  \\
  $t_{0,2}$ $(\rm{HJD}^{\prime})$ &...                   &...                  &7557.049 $\pm$ 0.064 \\
  $u_{0,2}$                       &...                   &...                  &0.027 $\pm$ 0.013   \\
  $\rho_{2}$ $(10^{-2})$          &...                   &...                  &6.417 $\pm$ 0.877    \\
  $q_{F,I}$ $(10^{-3})$           &...                   &...                  &2.760 $\pm$ 0.306    \\
  $f_s$                           &4.065 $\pm$ 0.536     &4.201 $\pm$ 0.522    &3.977 $\pm$ 0.622    \\
  $f_b$                           &-0.439 $\pm$ 0.536    &-0.575 $\pm$ 0.522   &-0.351 $\pm$ 0.622    \\
    \cline{1-4}
  $t_*$ $(\rm{days})$             &1.215 $\pm$ 0.124     &3.752 $\pm$ 0.394    &...                  \\
  $t_{\rm eff}$ $(\rm{days})$     &18.919 $\pm$ 0.445    &18.987 $\pm$ 0.415   &...                  \\
  \enddata
\label{tab:ulens}
\end{deluxetable}

\begin{deluxetable}{lcc}
\tablecolumns{3} \tablewidth{0pc}\tablecaption{\textsc{Physical
properties}}
\tablehead{\colhead{Quantity}&\colhead{$s<1$}&\colhead{$s>1$}}
\startdata
  $M_{\rm host}$ $[M_\sun]$      &$0.087_{-0.049}^{+0.092}$  &$0.022_{-0.009}^{+0.023}$ \\
  $M_{\rm planet}$ $[M_J]$       &$3.283_{-1.835}^{+3.468}$  &$0.090_{-0.037}^{+0.096}$\\
  $D_{\rm L}$ [kpc]              &$6.651_{-1.348}^{+0.948}$  &$7.481_{-0.708}^{+0.748}$ \\
  $a_\bot$ [au]                  &$0.342_{-0.085}^{+0.070}$  &$1.065_{-0.189}^{+0.192}$ \\
  \enddata
\label{tab:phys}
\end{deluxetable}

\acknowledgments 
Work by AG was supported by AST-1516842 from the US NSF.
IGS and AG were supported by JPL grant 1500811.
Work by C.H. was supported by the grant (2017R1A4A1015178) of 
National Research Foundation of Korea.
This research has made use of the KMTNet system operated by the Korea
Astronomy and Space Science Institute (KASI) and the data were obtained at
three host sites of CTIO in Chile, SAAO in South Africa, and SSO in
Australia.

\begin{thebibliography}{99}


\bibitem[Alard \& Lupton(1998)]{alard98} Alard, C. \& Lupton, R.H.,1998, \apj, 503, 325



\bibitem[Albrow et al.(2009)]{albrow09}Albrow, M.\ D., Horne, K., Bramich, D.\ M., et al.\ 2009, \mnras, 397, 2099







\bibitem[Batista et al.(2015)]{ob05169bat} Batista, V., Beaulieu, J.-P., Bennett, D.P., et al. 2015, \apj, 808, 170

\bibitem[Beaulieu et al.(2006)] {ob05390}Beaulieu, J.-P. Bennett, D.P., Fouqu\'e, P. et al. 2006, Nature, 439, 437






\bibitem[Bensby et al.(2013)]{bensby13} Bensby, T. Yee, J.C., Feltzing, S.\ et al.\ 2013, \aap, 549A, 147

\bibitem[Bessell \& Brett(1988)]{bb88} Bessell, M.S., \& Brett, J.M.\ 1988, \pasp, 100, 1134



















\bibitem[Dominik(1999)]{dominik99} Dominik, M. 1999, \aap, 349, 108



\bibitem[Gaudi(1998)]{gaudi98} Gaudi, B.S.\ 1998, \apj, 506, 533











\bibitem[Gould(1997)]{gould97} Gould, A. 1997, The Hollywood Strategy for Microlensing Detection of Planets, in Variables Stars and the Astrophysical Returns of the Microlensing Surveys. Eds. R. Ferlet, J.-P. Maillard and B. Raban. Gif-sur-Yvette, France : Editions Frontieres, p.125






\bibitem[Gould \& Gaucherel(1997)]{gg97} Gould, A. \& Gaucherel. 1997, \apj, 477, 580

\bibitem[Gould \& Horne(2013)]{gouldhorne} Gould, A. \& Horne, K. 2013, \apjl, 779, L28















\bibitem[Griest \& Safizadeh(1998)]{griest98} Griest, K.\ \& Safizadeh, N.\ 1998, \apj, 500, 37

\bibitem[Han \& Gould(1995)]{han95} Han, C. \& Gould, A.\ 1995, \apj, 447, 53


\bibitem[Han(2006)]{han06} Han, C.  2006, \apj, 638, 1080


\bibitem[Henderson et al.(2016)]{henderson16} Henderson, C.B., Poleski, R., Penny, M. et al. 2016 \pasp, 128, 124401


\bibitem[Hwang et al.(2018)]{ob170173} Hwang, K.-H., Udalski, A., Shvartzvald, Y. et al. 2018, \aj, 155, 20 





\bibitem[Jung et al.(2018)]{ob171522}Jung, Y.\ K., Udalski, A., Gould, A., et al.\ 2018 \aj, 155, 219



\bibitem[Kervella et al.(2004)]{kervella04} Kervella, P., Th{\'e}venin, F., Di Folco, E., \& S{\'e}gransan, D.\ 2004, \aap, 426, 297

\bibitem[Kim et al.(2016)]{kmtnet} Kim, S.-L., Lee, C.-U., Park, B.-G., et al.  2016, JKAS, 49, 37

\bibitem[Kim et al.(2018a)]{eventfinder} Kim, D.-J., Kim,  H.-W., Hwang, K.-H., et al., 2018a, \aj, 155, 76

\bibitem[Kim et al.(2018b)]{2016k2} Kim,  H.-W., Hwang, K.-H., Kim, D.-J., et al., 2018b, \aj, 155, 186

\bibitem[Kim et al.(2018c)]{2016eventfinder} Kim,  H.-W., Hwang, K.-H., Kim, D.-J., et al., 2018c, AAS submitted, arXiv:1804.03352






\bibitem[Nataf et al.(2013)]{nataf13} Nataf, D.M., Gould, A., Fouqu\'e, P. et al. 2013, \apj, 769, 88


\bibitem[Paczy\'nski(1986)]{pac86} Paczy\'nski, B.\ 1986, \apj, 304, 1





\bibitem[Poleski et al.(2014)]{ob08092} Poleski, R., Skowron, J., Udalski, A.,  et al.\ 2014, \apj, 755, 42

\bibitem[Poleski et al.(2017)]{mb12006} Poleski, R., Udalski, A.,  Bond, I.A., et al.\ 2017, \aap, 604A, 103


































\bibitem[Sumi et al.(2016)]{mb13605} Sumi, T., Udalski, A., Bennett, D.P., et al.\ 2016 \apj, 825, 112








\bibitem[Udalski et al.(2018)]{ob171434} Udalski, A., Ryu, Y.-H., Sajadian, S., et al.\ 2018, Acta Astron., 68, 1









\bibitem[Yoo et al.(2004)]{ob03262} Yoo, J., DePoy, D.L., Gal-Yam, A.\ et al.\ 2004, \apj, 603, 139






\end{thebibliography}
\end{document}